\documentclass[aps,pra,twocolumn,superscriptaddress]{revtex4-1}

\usepackage[T1]{fontenc}
\usepackage{graphicx,amsmath,amssymb,amsfonts,dsfont}
\usepackage{multirow,colortbl}
\usepackage{hyperref}

\def\tr{\mathrm{tr}}

\def\ket#1{|#1\rangle}
\def\braket#1#2{\langle#1|#2\rangle}
\def\ketbra#1{|#1\rangle\langle#1|}
\def\bra#1{\langle#1|}

\def\seb#1{#1}

\newcommand{\dd}{\mathrm{d}}

\newcommand{\calp}{\mathcal{P}}
\newcommand{\calu}{\mathcal{U}}

\newcommand{\id}{\mathds{1}}

\newcommand{\ba}{\begin{equation}}
\newcommand{\ea}{\end{equation}}
\renewcommand{\leq}{\leqslant}
\renewcommand{\geq}{\geqslant}

\makeatletter
\def\CT@@do@color{%
  \global\let\CT@do@color\relax
  \@tempdima\wd\z@
  \advance\@tempdima\@tempdimb
  \advance\@tempdima\@tempdimc
  \advance\@tempdimb0.9\tabcolsep
  \advance\@tempdimc\tabcolsep
  \advance\@tempdima2\tabcolsep
  \kern-\@tempdimb
  \leaders\vrule
  \hskip\@tempdima\@plus 1fill
  \kern-\@tempdimc
\hskip-\wd\z@ \@plus -1fill }
\makeatother

\begin{document}

\title{Unbounded sequence of observers exhibiting Einstein-Podolsky-Rosen steering}

\author{Akshata Shenoy H.}
\thanks{These authors contributed equally to this work.}
\author{S\'ebastien Designolle}
\thanks{These authors contributed equally to this work.}
\author{Flavien Hirsch}
\affiliation{D\'epartement de Physique Appliqu\'ee, Universit\'e de Gen\`eve, 1211 Gen\`eve, Switzerland}
\author{Ralph Silva}
\affiliation{D\'epartement de Physique Appliqu\'ee, Universit\'e de Gen\`eve, 1211 Gen\`eve, Switzerland}
\affiliation{Institute for Theoretical Physics, ETH Z\"urich, 8093 Z\"urich, Switzerland}
\author{Nicolas Gisin}
\author{Nicolas Brunner}
\affiliation{D\'epartement de Physique Appliqu\'ee, Universit\'e de Gen\`eve, 1211 Gen\`eve, Switzerland}

\date{\today}

\begin{abstract}
A sequential steering scenario is investigated, where multiple Bobs aim at demonstrating steering using successively the same half of an entangled quantum state. With isotropic entangled states of local dimension $d$, the number of Bobs that can steer Alice is found to be $N_\mathrm{Bob} \sim d / \log{d}$, thus leading to an arbitrary large number of successive instances of steering with independently chosen and unbiased inputs. This scaling is achieved when considering a general class of measurements along orthonormal bases, as well as complete sets of mutually unbiased bases. Moreover, we show that similar results can be obtained in an anonymous sequential scenario, where none of the Bobs know their position in the sequence. \seb{Finally, we briefly discuss the implication of our results for sequential tests of Bell nonlocality.}
\end{abstract}

\maketitle

\section{Introduction}

Distant parties sharing an entangled quantum state can establish strong correlations that admit no analog in classical physics \cite{Ein35,Bel64}. Formally, quantum correlations can be captured via different concepts, such as entanglement, EPR steering and Bell nonlocality. All of these are inequivalent concepts \cite{Wis07,Qui15}, forming a hierarchy of quantum correlations where entanglement is the weakest form and Bell nonlocality the strongest; see, e.g., \cite{Bru14,Cav17} for recent reviews.

The standard scenario for discussing these phenomena involves two ingredients: (i) a source distributing an entangled state to distant parties, and (ii) sets of local measurements performed by the parties. In recent years however, it was suggested to add another ingredient to this picture, namely, quantum channels. This leads to a scenario where several quantum measurements can be performed sequentially on the same quantum subsystem.

This idea was first proposed in order to reveal the ``hidden nonlocality'' of certain entangled states admitting a local model \cite{Pop95,Gis96,Hir13}.
This amounts to each party first applying a local filter (i.e., a quantum channel) on their subsystem, and then performing a standard Bell test.
However, there exist certain entangled states which feature no hidden nonlocality \cite{Hir16}, which motivates the study of more sophisticated sequential Bell tests \cite{Gal14}.

More recently, Ref.~\cite{Sil15} developed a different use of channels in the context of Bell tests.
Instead of having each party performing a sequence of \seb{possibly correlated} measurements \seb{(e.g. using feed-forward)}, there is now a sequence of parties, each of which performs a measurement on the same half of an entangled state successively \seb{and independently}. Hence each party receives a quantum system and performs a non-destructive measurement on it (represented by a quantum channel), and passes the system over to the next party in the sequence. Importantly, it is now the full description of the measurement that matters, i.e., not only the positive operator-valued measure (POVM) elements, but also the Kraus operators which also characterize the post-measurement state.

Specifically, Ref.~\cite{Sil15} showed that the Bell nonlocality of a two-qubit maximally entangled pair could be shared by Alice (on one side), and two Bobs (on the other side). Moreover, they showed that an arbitrary long sequence of Bobs can establish nonlocal correlations with Alice, given that their measurement inputs are judiciously biased. This motivated further work exploring the potential of these ideas for randomness generation \cite{Cur17} and their classical communication cost \cite{Tav18}, and led to experimental demonstrations \cite{Sch17,Hu16}. More recently, these ideas were extended to \seb{other types of quantum correlations \cite{Ber18,Sah18,Sas18}.} Specifically, Ref.~\cite{Sas18} showed that three Bobs, each performing three different measurements, could steer Alice using a two-qubit maximally entangled pair.

In the present work, we consider the general scenario of sequential steering, featuring an arbitrary number of Bobs steering the state of Alice. We show that for symmetric entangled states of arbitrary dimension $d \times d$, \seb{and measurements of L\"uders form,} the problem can essentially be completely solved. Considering first a situation where the $i$th Bob can choose an unsharpness parameter $\eta_i$ and perform any measurement in $\calp_{\eta_i}$ (see Eq.~\eqref{eqn:peta}), we show that the number of Bobs that can steer Alice grows as $N_\mathrm{Bob} \sim d / \log{d}$. 
Hence an unbounded number of Bobs can steer Alice, using independently chosen and unbiased measurement settings. Then we show that the same result can be obtained under the restriction that each Bob performs only a finite set of measurements, specifically, w.r.t.~complete sets of mutually unbiased bases. We also show that these results can be extended to the scenario in which the Bobs do not know their position in the sequence.
\seb{
  Next, we briefly discuss the implications of our results for sequential tests of Bell nonlocality. In particular, we find that for symmetric entangled states of two qubits, at most two Bobs can violate a Bell inequality, considering an arbitrary number of independently chosen measurement settings.
}
Finally, we conclude with a list of open questions.

In this paper, we use $\calp_\eta$ to denote the set of measurements obtained by mixing projective measurements with white noise, namely,
\begin{equation}
  \label{eqn:peta}
  \calp_\eta=\left\{\left(\eta\ketbra{\phi_i}+(1-\eta)\frac{\id}{d}\right)_{i=1\ldots d}\bigg||\braket{\phi_i}{\phi_j}|^2=\delta_i^j\right\}.
\end{equation}
For $\eta=1$, $\calp_1$ contains all projective measurements whereas, for $\eta=0$, $\calp_0$ is reduced to white noise only.

\section{Preliminaries}

\subsection{Steering}

We start by recalling the usual (i.e., non-sequential) steering scenario.
Alice and Bob share an entangled state quantum state $\rho$. Let Alice be allowed to fully characterize her part of the shared state, while Bob performs measurements represented by POVMs $B_{b|y}$ with inputs $y$ and corresponding outcomes $b$.
Once Bob has performed his measurement, Alice is left with an sub-normalized assemblage of states
\begin{equation}
  \sigma_{b|y}=\tr_B(\id\otimes B_{b|y})\rho \,.
\end{equation}
such that the no-signalling condition is satisfied, i.e., $\sum_b\sigma_{b|y}=\rho_A = \tr_B (\rho)$ for all $y$. If the assemblage can be explained by a local hidden state (LHS) model, namely,
\begin{equation}
  \sigma_{b|y}=\int\pi(\lambda)p_B(b|y,\lambda)\sigma_\lambda\dd\lambda,
\end{equation}
\seb{for any variable $\lambda$, distributed with density $\pi(\lambda)$, and for any local response distribution $p_B (b|y,\lambda)$,} then the state is said to be unsteerable from Bob to Alice with respect to the chosen set of measurements $B_{b|y}$. On the contrary, when such model can be proven not to exist, we say that steering from Bob to Alice is demonstrated, as Bob can remotely steer Alice's state in a way that admits no local explanation.

\subsection{Symmetric states}
\label{subsec:sym}

In our work, we will focus on two classes of bipartite entangled state featuring a strong degree of symmetry, namely, Werner and isotropic states. Werner states satisfy, for all unitary $U$, the invariance property \cite{Wer89}
\begin{equation}
  \rho_W=U\otimes U\rho_WU^\dag\otimes U^\dag \,.
\end{equation}
and can be specified, once the local dimension is fixed to $d$, with a single parameter $0\leq p\leq 1$, namely,
\begin{equation}
  \rho_W(p)=\frac{1}{d(d-1)}\left(\frac{d-1+p}{d}\id-pV\right),
\end{equation}
where $V=\sum_{i,j}\ket{ij}\bra{ji}$ is the swap operator.

Similarly, isotropic states verify
\begin{equation}
  \rho_\mathrm{iso}=U^*\otimes U\rho_\mathrm{iso}(U^*)^\dag\otimes U^\dag
\end{equation}
and can be parametrized by $0\leq p\leq1$ through
\begin{equation}
  \rho_\mathrm{iso}(p)=p\ketbra{\phi^+}+(1-p)\frac{\id}{d^2},
\end{equation}
where $\ket{\phi^+}=\frac{1}{\sqrt{d}}\sum_{i=0}^{d-1}\ket{i}\otimes\ket{i}$ is a maximally entangled state of dimension $d \times d$.

Both Werner and isotropic states are known to be separable if and only if $p\leq1/(d+1)$.
Following Ref.~\cite{Wis07}, we also know the exact characterization of the steerability of these classes of states when using all projective measurements.
Specifically, we are interested in the threshold $p_\mathrm{steer}$ above which Bob can exhibit steering by means of all (noiseless) projective measurement and below which a LHS model exists. We have that
\begin{align}
  p_\mathrm{steer}^{W}(d) = \frac{d-1}{d}  \quad  \text{and} \quad  p_\mathrm{steer}^{\text{iso}}(d) = \frac{\sum_{i=2}^d \frac{1}{i}}{d-1}.
\label{eqn:psteer}
\end{align}

\section{Steering with multiple Bobs}

\begin{figure}[t!]
  \centering
  \includegraphics[height=2cm]{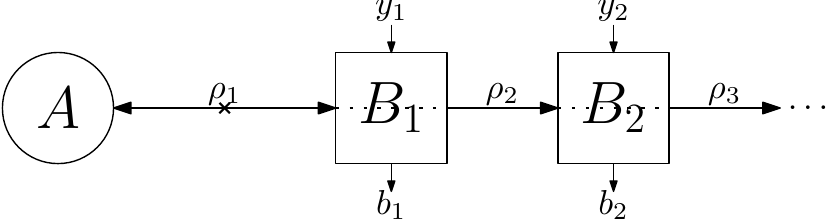}
  \caption{Sequential steering scenario: multiple Bobs aim at steering Alice. Note that each state $\rho_i$ is a shared state between Alice and the $i$-th Bob $B_i$.}
  \label{fig:bobs}
\end{figure}

We consider a sequential steering scenario, which features an arbitrary number of Bobs, each of them trying to steer the state of Alice (see Fig.~\ref{fig:bobs}). That is, each Bob $B_i$ aims at generating an assemblage for Alice that demonstrates steering.

Formally, $B_i$ shares a state $\rho_i$ (acting on a Hilbert space of dimension $d \times d$) with Alice. Upon receiving an input $y_i$, $B_i$ applies a quantum measurement on his subsystem. This produces (i) a measurement output $b_i$, and (ii) an output (or postmeasurement) state $\rho_{i+1}$ that will be shared between Alice and the next Bob $B_{i+1}.$
Importantly, we assume that all inputs $y_i$ are chosen uniformly at random, and are uncorrelated to each other.

The channels describing the quantum operation implemented by each Bob are represented by a set of Kraus operators
$\{K_{b_i|y_i}^{(i)}\}_{b_i}$, for each input $y_i$. The POVM describing $B_i$'s measurement is $B_{b_i|y_i}^{(i)} = K_{b_i|y_i}^{(i)\dag}K_{b_i|y_i}^{(i)}$, while the output (or postmeasurement) state is given by
\begin{equation}
  \rho_{i+1} = \sum_{y_i,b_i}(\id\otimes K_{b_i|y_i}^{(i)})\rho_i(\id\otimes K_{b_i|y_i}^{(i)\dag}) \,.
\end{equation}

We will focus on $d$-output measurements in $\calp_\eta$. These can also be viewed as noisy implementations of projective measurements.
The reason for using this kind of measurement is quite intuitive.
If we use projective measurements, only the first Bob can exhibit steering since the disturbance he would introduce would prevent all the following Bobs from doing so.
Unsharpening the measurements with a parameter $\eta$ gives the ability to tune the trade-off between disturbance and information gain.

When the $i$th Bob chooses an unsharpness parameter $\eta_i$, he can perform any measurement in $\calp_{\eta_i}$, namely, ${B_{b_i|y_i}^{(i)}=\eta_i\ketbra{\phi_{b_i|y_i}^{(i)}}+(1-\eta_i)\id/d}$. A natural choice for the Kraus operators, usually named after L\"uders \cite{Bus86}, is ${K_{b_i|y_i}^{(i)}=\sqrt{B_{b_i|y_i}^{(i)}}}$. Here we get $K_{b_i|y_i}^{(i)}$ of the form
\begin{equation}
  \label{eqn:kraus}
\left(\sqrt{\frac{1+(d-1)\eta_i}{d}}-\sqrt{\frac{1-\eta_i}{d}}\right)\ketbra{\phi_{b_i|y_i}^{(i)}}+\sqrt{\frac{1-\eta_i}{d}}\id.
\end{equation}
In the qubit case, the measurements we are interested in are of the form
\begin{equation}
  B_{\pm|\hat{y}}=\eta\ketbra{\pm_{\hat{y}}}+(1-\eta)\frac{\id}{2}=\frac{\id\pm\eta\hat{y}\cdot\vec{\sigma}}{2},
\end{equation}
where $\hat{y}$ is the direction of the measurement in the Bloch sphere.
Then Eq.~\eqref{eqn:kraus} reduces to
\begin{equation}
  K_{\pm|\hat{y}}=\frac{1}{\sqrt{2}}(\sqrt{1+\eta}\ketbra{+_{\hat{y}}}+\sqrt{1-\eta}\ketbra{-_{\hat{y}}}).
\end{equation}

As we are dealing with measurements based on orthonormal bases, it is convenient to introduce the unitary matrices $U_{b_i|y_i}^{(i)}$ defined by $U_{y_i}^{(i)}\ket{b_i}=\ket{\phi_{b_i|y_i}^{(i)}}$, where $\{\ket{b_i}\}_{b_i}$ is the computational basis.
Then it is straightforward to see that ${K_{b_i|y_i}^{(i)}=U_{y_i}^{(i)}K_{b_i}^{(i)}U_{y_i}^{(i)\dag}}$, with $K_{b_i}^{(i)}$ defined as in Eq.~\eqref{eqn:kraus} but with $\ketbra{b_i}$ instead of $\ketbra{\phi_{b_i|y_i}^{(i)}}$.

Let us comment on the use of L\"uders measurement. Clearly, the sequential steering scenario requires measurement channels that minimize disturbance (of the post-measurement state) given a certain information gain, or vice versa. Intuitively, L\"uders measurements are a good choice. More formally, this can be verified for the qubit case, using the figures of merit introduced by in Ref.~\cite{Sil15}, where disturbance is quantified via the fidelity $F$ of the post-measurement state with respect to the input state, while information gain is characterized by the strength of the POVM $G$. For L\"uders measurements, \seb{it was shown in Ref.~\cite{Mal16} that} $F = \sqrt{1-\eta^2} $ and $G = \eta$, thus clearly saturating the inequality $F^2 + G^2 \leq 1$ derived in \cite{Sil15}. This provides good evidence that these measurements are optimal in the present context.

\subsection{All measurements}

Let us consider the case where each Bob is given the choice to perform any measurement of the form of Eq.~\eqref{eqn:kraus} and chooses uniformly among them. As we assume that all Bobs are independent, $B_{i+1}$ does not know the input and output of $B_i$, so that the state he receives is obtained by averaging over all possible channels, namely,
\begin{equation}
  \label{eqn:proj}
  \rho_{i+1}=\int\dd U\sum_{b_i}(\id\otimes UK_{b_i}^{(i)}U^\dag)\rho_i(\id\otimes U^\dag K_{b_i}^{(i)\dag}U),
\end{equation}
where $\dd U$ is the uniform (Haar) measure on $d\times d$ unitaries. From this we can immediately see that if $\rho_i$ is invariant over conjugation by either $U\otimes U$ or $U^*\otimes U$, then so is $\rho_{i+1}$.
Therefore, if the initial state is either a Werner state or an isotropic state, then the whole sequence of states $\rho_i$ will inherit from this symmetry and thus stay in the same class.
Moreover, following Ref.~\cite{Eme05}, we can use the invariance of the integral \eqref{eqn:proj} to express the parameter $p_{i+1}$ as a function of $p_i$ and $\eta_i$
\begin{align}
  p_{i+1}&=\sum_{b_i}\frac{\left(\tr\,K_{b_i}^{(i)}\right)^2-\frac1d\tr\,\left(K_{b_i}^{(i)}\right)^2}{d^2-1}\\
  &=\frac{\eta_i+(1-\eta_i)(d-1)+2\sqrt{1-\eta_i}\sqrt{1+(d-1)\eta_i}}{d+1}p_i.\label{eqn:rec}
\end{align}
Interestingly, since this expression only relies on symmetries of the integral in Eq.~\eqref{eqn:proj}, it holds for both Werner and isotropic states.

Now we are able to find out how many Bobs can demonstrate steering.
Because of the structure of both Werner and isotropic states, one can transfer the unsharpness of the measurement into the state, namely,
\begin{equation}
  \tr_B(\id\otimes B_b^\eta)\rho_\mathrm{iso}(p)=\tr_B(\id\otimes B_b)\rho_\mathrm{iso}(\eta p),
\end{equation}
and similarly for Werner states.
Therefore, by using the results of Ref.~\cite{Wis07} recalled in Eq.~\eqref{eqn:psteer}, $B_i$ can exhibit steering if and only if
\begin{equation}
  \eta_i p_i > p_\mathrm{steer}.
  \label{eqn:steer}
\end{equation}

\begin{figure}[h]
  \centering
  \includegraphics[width=8cm]{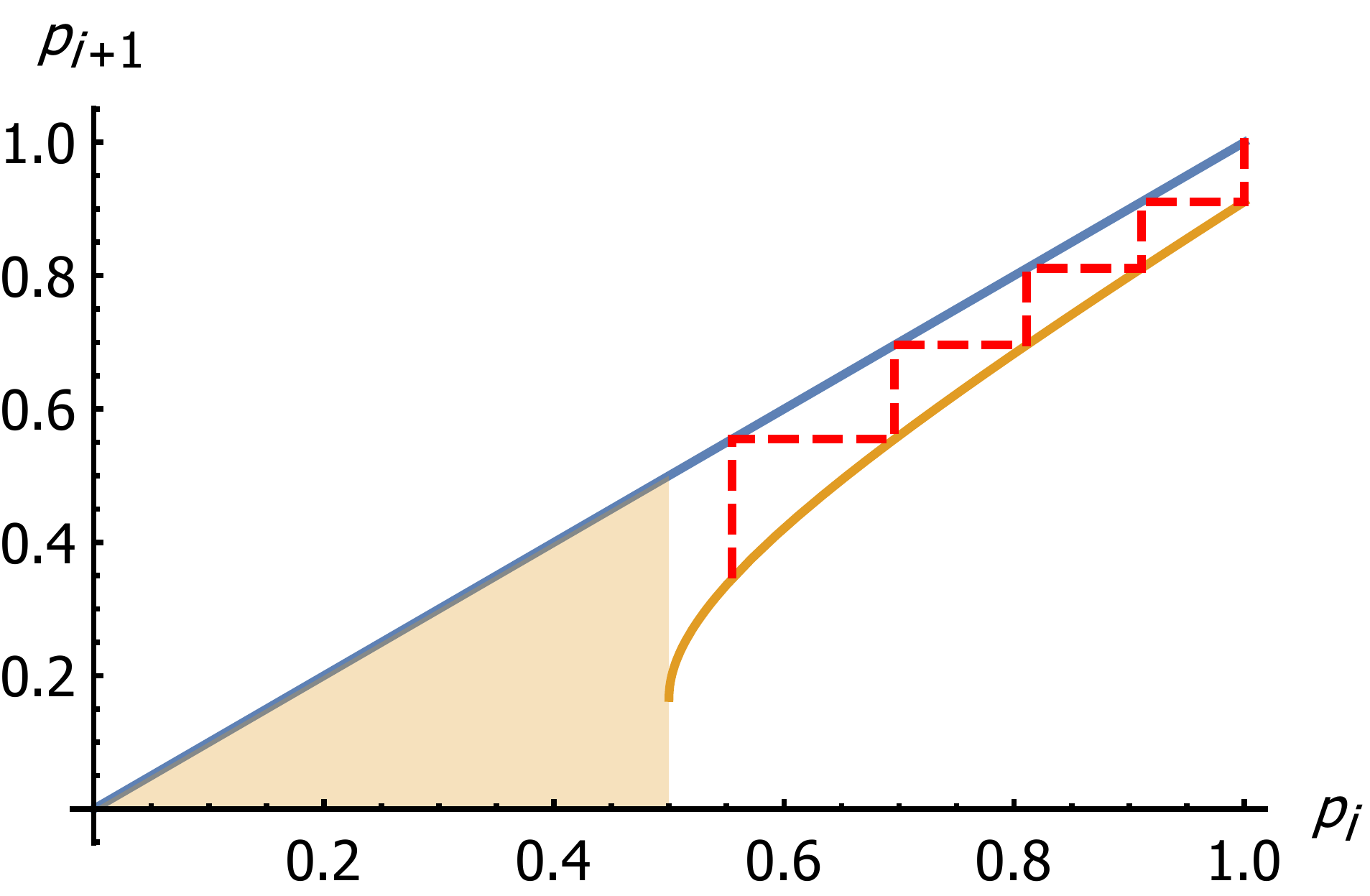}
  \caption{Graphical representation of the number of Bobs that are able to exhibit steering in the limit case in which each of them just saturates the steering criterion, here for the case of qubits $d=2$. In the shaded area, the noise is too strong for steering to be demonstrated. \seb{As the dashed line can only do five steps before falling in this area, we see that (at most)} five Bobs can exhibit steering.}
  \label{fig:ill}
\end{figure}

It is insightful to explore graphically the behavior of the sequence $\{p_i\}_i$ that saturates the steering criterion \eqref{eqn:steer} for every Bob. It is fully characterized by the initial value $p_1=1$ and the recursive equation \eqref{eqn:rec} into which we plug $\eta_i=p_\mathrm{steer}/p_i$.
In the following we will refer to the latter as the saturating function.
In Fig.~\ref{fig:ill}, we plot it for $d=2$.

Then the construction of the sequence has a clear geometrical meaning.
Moreover, the maximal number of Bobs that can exhibit steering is reached whenever the sequence goes below $p_\mathrm{steer}$.
In dimension two, Werner and isotropic states coincide and enable up to five consecutive Bobs to demonstrate steering of Alice.
\seb{
  Note that it was conjectured in Ref.~\cite{Sas18} that $N$ Bobs can exhibit steering with respect to Alice when steering is probed through an $N$-settings steering inequality, in the case $d=2$.
  From our results, which are admittedly restricted to the maximally entangled state together with specific measurements, it seems that this conjecture is not true.
}

\begin{figure}[b!]
  \centering
  \includegraphics[width=8cm]{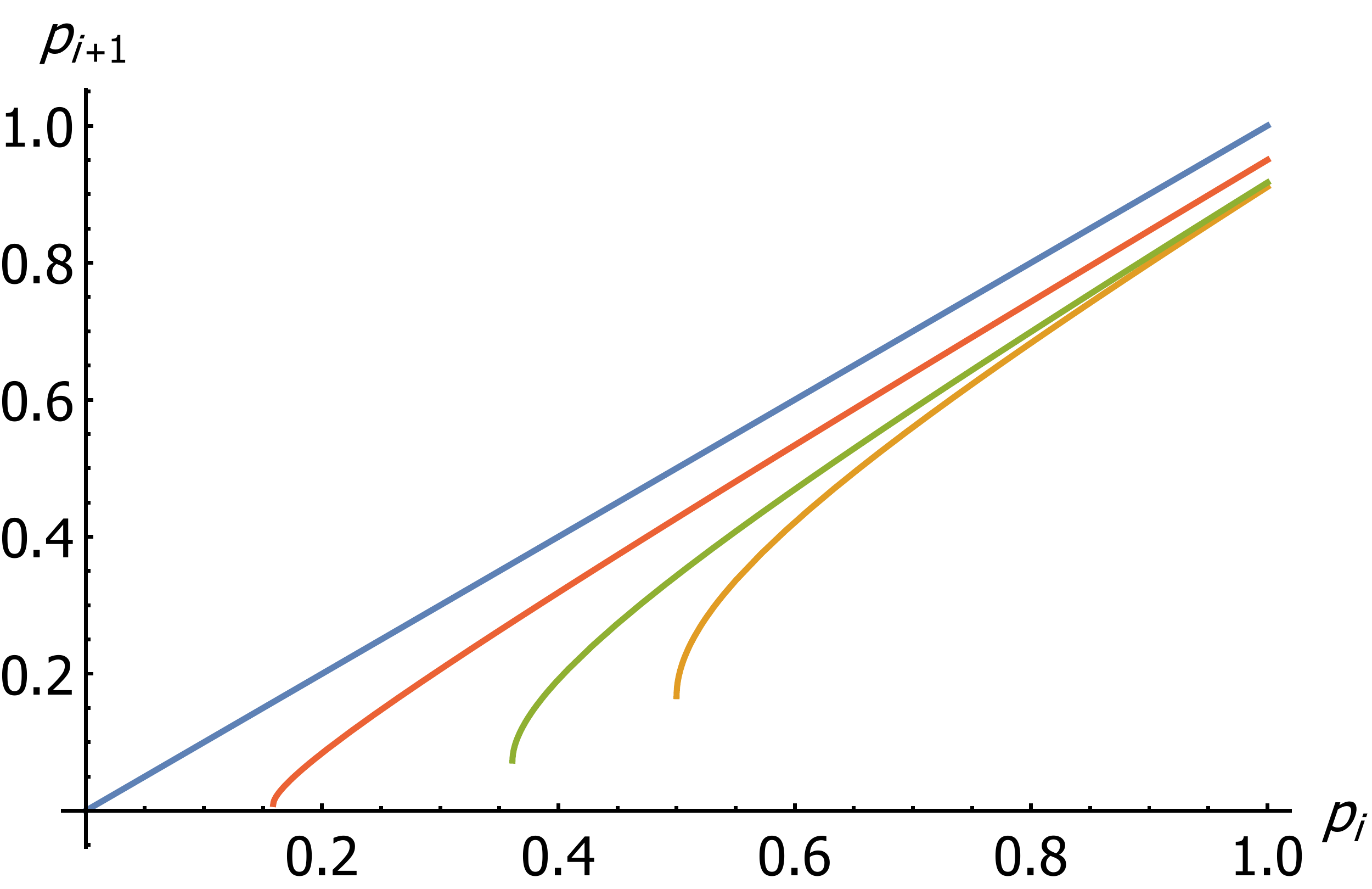}
  \caption{Similar curves as in Fig.~\ref{fig:ill}, here for isotropic states in dimensions 2, 4, and 16 (from bottom to top). The noise thresholds $p_\mathrm{steer}^{\text{iso}}(d)$ are approximately 0.50, 0.36, and 0.16 (respectively), and 5, 6, and 13 Bobs can demonstrate steering (respectively). These numbers of Bobs correspond to the last $i$ such that $p_i \geq p_\mathrm{steer}^{\text{iso}}(d)$.}
  \label{fig:dim}
\end{figure}

In higher dimensions, for Werner states, as $p_\mathrm{steer}^{W}$ increases with $d$, fewer and fewer Bobs can exhibit steering, so that two Bobs can do so for $d=3$, and, from $d=4$, only one Bob can do so.
With isotropic states, the behavior is different as $p_\mathrm{steer}^{\text{iso}}(d)$ decreases with $d$, and tends to 0 for large $d$. For increasing $d$, the saturating function therefore gets closer to maintaining the parameter $p_{i+1}=p_i$ for the next Bob, so that the corresponding sequence $\{p_i\}_i$ decreases slower.
We illustrate this effect in Fig.~\ref{fig:dim}.
Therefore, the number of Bobs, $N_\mathrm{Bob}$, that can exhibit steering increases with the dimension $d$.
More precisely, by lower and upper bounding the saturating function by suitable linear functions, we can show, by means of a symbolic computation software, that
\begin{equation}
  \label{eqn:nproj}
  N_\mathrm{Bob}\sim\frac{d}{\log d},
\end{equation}
where $\log$ is the natural logarithm.
This equivalent is a lower bound for the actual value of $N_\mathrm{Bob}$.
For $d\leq150$, it underestimates $N_\mathrm{Bob}$ by roughly a factor of 2.

Importantly, note that in this construction each Bob must know his respective position in the sequence, as each one performs a measurement with a well chosen strength $\eta_i=p_\mathrm{steer}/p_i$. Nevertheless, we will see in Sec.~\ref{anonymous} that this condition can be dispensed with, while keeping the scaling of Eq.~\eqref{eqn:nproj}.

Also, note that the above construction is optimal for symmetric states, when considering L\"uders measurements of the form of Eq.~\eqref{eqn:kraus}. Indeed, as the states are symmetric there can be no preferred measurement direction, and hence the best possibility consists in considering all possible measurements directions on an equal footing.



\subsection{Finite sets of measurements}

While the case of all possible measurements is interesting conceptually, it is also relevant to ask (e.g., from a practical point of view) whether similar results can be obtained when each Bob performs only a finite set of measurements. A natural idea here is to use 2-designs in order to replace the integral \eqref{eqn:proj} by a sum \cite{DiM14}, hence mapping the previous case to that of a finite set of measurements.
Intuitively, 2-designs are simply finite sets over which the average reproduces the average over the total (infinite) set.

We consider first the case of a unitary 2-design $\calu$ \cite{Dan09}.
If $B_i$ performs measurements in the bases corresponding to the elements of $\calu$, then
\begin{align}
  \rho_{i+1}&=\sum_{U\in\calu}\sum_{b_i}(\id\otimes UK_{b_i}^{(i)}U^\dag)\rho_i(\id\otimes UK_{b_i}^{(i)}U^\dag)\\
  &=\int\dd U\sum_{b_i}(\id\otimes UK_{b_i}^{(i)}U^\dag)\rho_i(\id\otimes U^\dag K_{b_i}^{(i)\dag}U),
\end{align}
which has the same invariance property as Eq.~\eqref{eqn:proj}. Thus, if all Bobs use the elements of $\calu$ as their measurements bases, the evolution of both Werner and isotropic states would be exactly the same as in the previous case. Despite the simplicity of the argument, applying it to our problem is not straightforward. First, only few exact constructions of unitary-designs are known. To the best of our knowledge, it is only known when $d=2^n$, in which case the Clifford group is a unitary 2-design \cite{DiV02}. More importantly, even with this example, the steering properties have not been characterized, that is, we do not know the threshold values of $p_{\text{steer}}$ for symmetric states and these measurements.


To overcome this last problem, we turn to a different kind of 2-designs, namely, complex projective 2-designs.
The difference with above only concerns the set in which  objects are embedded. Instead of unitaries we now have complex vectors.

Complete sets of mutually unbiased bases (CSMUB) are sets of $d+1$ bases in dimension $d$ such that any pair $\{\ket{\varphi_j}\}_j,\{\ket{\psi_k}\}_k$ satisfies $|\braket{\varphi_j}{\psi_k}|=1/\sqrt{d}$ for all $j$ and $k$.
They have numerous applications and have therefore been widely studied.
In particular, CSMUB are complex projective 2-designs \cite{Kla05}, so that, if $B_i$ performs measurements in these bases, then
\begin{align}
  \rho_{i+1}&=\frac{1}{d+1}\sum_{y_i}\sum_{b_i}(\id\otimes K_{b_i|y_i}^{(i)})\rho_i(\id\otimes K_{b_i|y_i}^{(i)\dag})\\
  &=d\int\dd\phi[\id\otimes K^{(i)}(\phi)]\rho_i[\id\otimes K^{(i)\dag}(\phi)]\\
  &=d\int\dd U(\id\otimes UK_{1}^{(i)}U^\dag)\rho_i(\id\otimes U^\dag K_{1}^{(i)\dag}U)\\
  &=\int\dd U\sum_{b_i}(\id\otimes UK_{b_i}^{(i)}U^\dag)\rho_i(\id\otimes U^\dag K_{b_i}^{(i)\dag}U),
\end{align}
which has again the desired invariance property. The last equality is obtained by replacing the factor $d$ by the sum over $b_i$, then remarking that in the integral the index $b_i$ can be used to address any other column of $U$ (since they play similar roles), and eventually to permute the sum and the integral to recover exactly Eq.~\eqref{eqn:proj}.

Interestingly, the steering properties of CSMUB have already been characterized \cite{Des18}.
From this work, and using the general connection between steering and measurement incompatibility \cite{Qui14,Uol14,Uol15}, it follows that, for isotropic states measured with (noiseless) CSMUB, we have
\begin{equation}
  p_\mathrm{steer}^{\text{iso}}(d)\leq\frac{\frac{\sqrt{d}}{d+1}+1}{\sqrt{d}+1}.
\end{equation}
Thus we can follow the same construction as in the previous section to show that an unbounded number of Bobs can exhibit steering, namely,
\begin{equation}
  \label{eqn:nmub}
  N_\mathrm{Bob}^\mathrm{MUB}\sim\frac{d}{\log d}.
\end{equation}

\seb{
  Note that the above analysis cannot be directly applied to Werner states. As these are never pure states (even for $p=1$) in dimension $d\geq 3$, their steering properties can no longer be related to the incompatibility of MUB.
}


\section{Anonymous sequential steering}\label{anonymous}

The results given in the previous sections were obtained using the fact that each Bob knows his position in the sequence. It is therefore natural to ask whether similar results could also be obtained when none of the Bobs are aware of their position. That is, each Bob must now act independently of all other Bobs. Here we show that in this \emph{anonymous} scenario, the number of Bobs that can demonstrate steering remains unbounded.

\begin{figure}[t!]
  \centering
  \includegraphics[width=8cm]{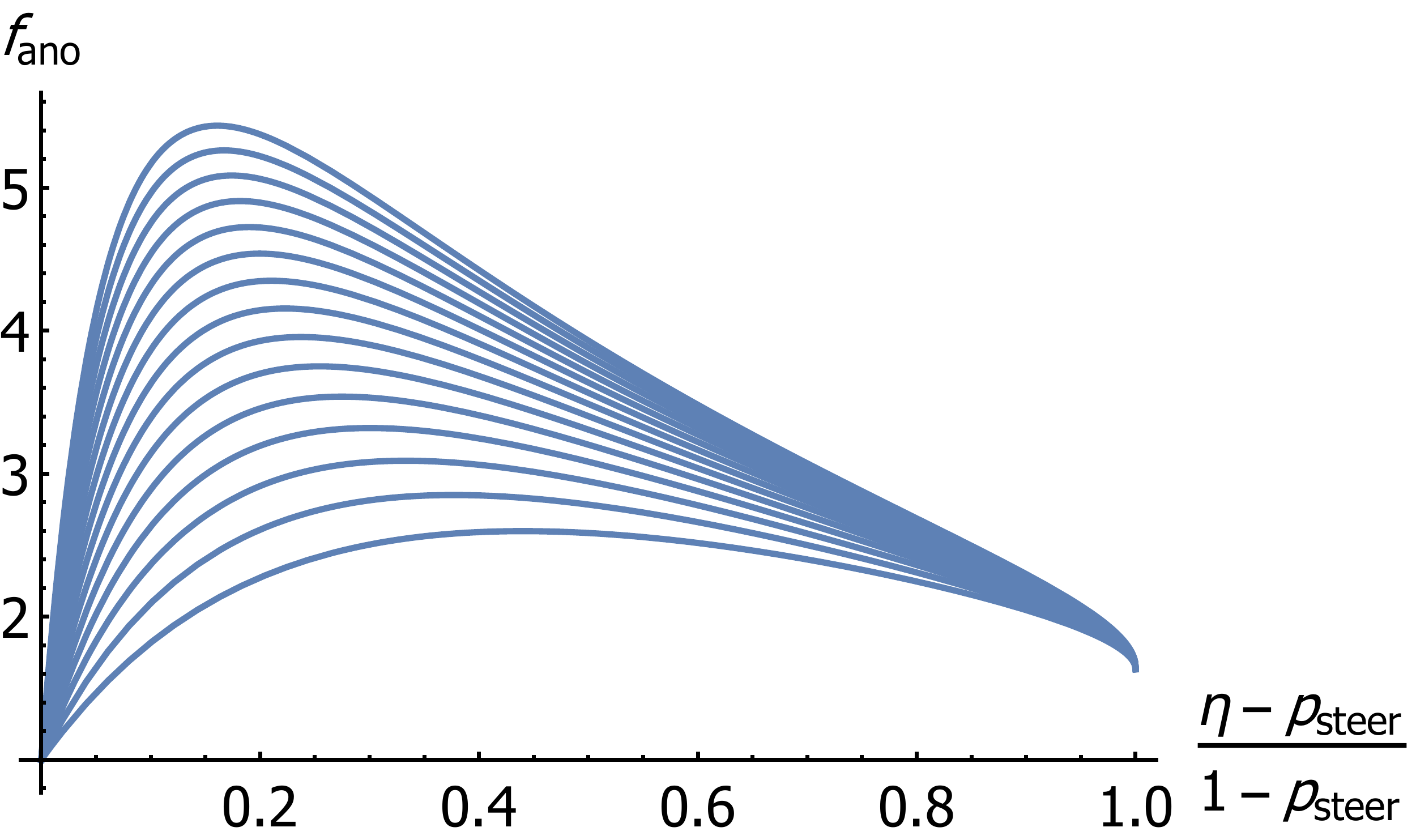}
  \caption{Behaviour of the function $f_\mathrm{ano}$ for $d=2\ldots16$. The horizontal axis has been rescaled to allow for a meaningful comparison of the curves.}
  \label{fig:anod}
\end{figure}

As each Bob does not know how many other Bobs have previously tried to steer Alice (and how many could still follow after him), the only option is that all Bobs perform measurements of the same strength, i.e., set the noise parameters to a constant, i.e., $\eta_i = \eta$ for all $i$. Then from Eq.~\eqref{eqn:rec} we get
\begin{equation}
  p_i=\left(\frac{\eta+(1-\eta)(d-1)+2\sqrt{1-\eta}\sqrt{1+(d-1)\eta}}{d+1}\right)^{i-1}.
\end{equation}
We call $r(\eta,d)$ the ratio of this geometric sequence.
From Eq.~\eqref{eqn:steer}, and considering the case of an isotropic state, we get a condition for $B_i$ to be able to steer Alice, namely,
\begin{equation}
  i<f_\mathrm{ano}(\eta,d)=1+\frac{\log{(p_{\text{steer}}^{\text{iso}}(d)}/\eta)}{\log r(\eta,d)}.
\end{equation}
Clearly, the function $f_\mathrm{ano}$, plotted in Fig.~\ref{fig:anod} for ${d=2\ldots16}$, quantifies the number of Bobs that can exhibit steering, considering here all measurements.

The behavior of this function can be understood in intuitive terms. When $\eta$ is just above $p_\mathrm{steer}^{\text{iso}}(d)$, only the first Bob can exhibit steering. In order to get more Bobs to steer, one needs to increase $\eta$ to compensate for the decrease of the visibility of the state through the sequence. However, at some point, when $\eta$ is too large, the disturbance introduced by each Bob becomes so large that it prevents the following Bobs from steering. Therefore there is an optimal value of $\eta$ at which $f_\mathrm{ano}$ is maximal, and thus $N_\mathrm{Bob}$ as well.


For example, taking $\eta=2p_{\mathrm{steer}}^{\text{iso}}(d)[1-1/(4\log d)]$, we get the lower bound
\begin{equation}
  \label{eqn:nproj2}
  N_\mathrm{Bob}\geq u_d\sim\frac{\log2}{2}\cdot\frac{d}{\log d}.
\end{equation}
Hence we see that it is again possible for an unbounded number of Bobs to steer Alice.
\seb{
  Note that, with Werner states, already in dimension three, no more than one Bob can steer Alice in this anonymous scenario.
}

With CSMUB, the same analysis applies, leading to
\begin{equation}
  \label{eqn:nmub2}
  N_\mathrm{Bob}^\mathrm{MUB}\geq v_d\sim\frac{\log2}{2}\sqrt{d}.
\end{equation}
One might be surprised by the difference of scaling between Eqs \eqref{eqn:nproj2} and \eqref{eqn:nmub2}.
Let us comment on that point. Previously, in the non-anonymous case, the same scaling was obtained for all measurements and CSMUB (see Eqs \eqref{eqn:nproj} and \eqref{eqn:nmub}), as the proofs relied only on the fact that $p_{\text{steer}}^{\text{iso}}$ tends to 0 when $d$ goes to infinity.
In the scenario we are now considering, as fewer assumptions are needed, the scaling depends on the precise behavior of $p_\mathrm{steer}^{\text{iso}}(d)$. For the case of all measurements we have $p_\mathrm{steer}^{\text{iso}}(d)\sim \log d/d$ whereas $p_\mathrm{steer}^{\text{iso}}(d)\lesssim 1/\sqrt{d}$ for CSMUB.

\section{Implications for nonlocality}

\seb{
In this section, we briefly discuss the consequences of our results for Bell nonlocality.
As there cannot be Bell nonlocality without steering, we have immediately an \emph{upper bound} on the number of Bobs $N_\mathrm{Bob}^\mathrm{NL}$ that can violate a Bell inequality with Alice.
This upper bound is restricted to our framework, namely, symmetric states (Werner and isotropic, see Sec.~\ref{subsec:sym}) together with L\"uders measurements (see Eq.~\eqref{eqn:kraus}) of measurements in $\calp_\eta$ (see Eq.~\eqref{eqn:peta}).}

\seb{For isotropic states, the upper bounds are Eqs \eqref{eqn:nproj} and \eqref{eqn:nmub} in the standard scenario and Eqs \eqref{eqn:nproj2} and \eqref{eqn:nmub2} in the anonymous scenario, with infinite and finite sets of measurements respectively.
In dimension two, we can provide stronger results, as bounds for the Bell nonlocality of the isotropic state are known, namely, it is local for $p \lesssim 0.6829$ \cite{Hir17}, and nonlocal for $p \gtrsim 0.7012$ \cite{Bri16}.
Using our construction, we find that at most two consecutive Bobs can exhibit Bell inequality violation with Alice. This is possible when using the simple CHSH Bell inequality \cite{Sil15}, with only two measurement settings per party. Hence it seems that even adding more measurements for Alice and the Bobs, even considering infinite sets of measurements, does not help. }

\seb{For Werner states, with infinitely many measurements and each Bob knowing his position in the sequence, we get the upper bounds $N_\mathrm{Bob}^\mathrm{NL}\leq5$, $N_\mathrm{Bob}^\mathrm{NL}\leq2$, and $N_\mathrm{Bob}^\mathrm{NL}=1$ for $d=2$, $d=3$, and $d\geq4$ respectively. Note that for $d\geq 3$ it is not known whether the Werner state can violate a Bell inequality, even for $p=1$.
In the anonymous scenario, no more than one Bob can demonstrate nonlocality in any dimension.
For finite sets of measurements, our approach does not give any bound for Werner states.}

\section{Conclusion}

We discussed a scenario of sequential steering, where multiple Bobs aim at steering Alice. For symmetric entangled states (of dimension $d \times d$) and L\"uders measurements, the problem can be solved in general. We showed that the number of Bobs that can steer Alice grows as $N_\mathrm{Bob} \sim d / \log{d}$, considering both the cases of essentially all possible measurements as well as complete sets of MUB. This shows that an unbounded number of Bobs can demonstrate steering, while using independently chosen and unbiased inputs. Moreover, we showed that these conclusions also hold in a scenario where each Bob does not know his position in the sequence. \seb{Finally, we discussed the implication of our results for sequential tests of Bell nonlocality.}

It would be interesting to further investigate the sequential steering scenario. In particular, a natural question is whether the scaling we found for $N_\mathrm{Bob}$ is optimal or if it can be improved. For symmetric states, we conjecture that our construction is essentially optimal. One could consider measurements that are not of the L\"uders form, and more general POVMs. Nevertheless, we believe that none of these would help. Alternatively, one could consider other entangled states. However, solving the problem in this case is likely to be more challenging, due to the reduced symmetry.

\subsection*{Acknowledgments}

We acknowledge financial support from the Swiss national science foundation (Starting grant DIAQ and NCCR-QSIT).
Akshata Shenoy H is grateful to the Federal Commission for Scholarships, Switzerland for the Swiss Government Excellence Fellowship.



\begin{thebibliography}{99}

  \bibitem{Ein35} A.~Einstein, B.~Podolsky, and N.~Rosen, {\it Can quantum-mechanical description of physical reality be considered complete?}, Phys.~Rev.~{\bf 47}, 777 (1935).
  \bibitem{Bel64} J.S.~Bell, {\it On the Einstein-Podolsky-Rosen paradox}, Physics {\bf 1}, 195-200 (1964).
  \bibitem{Wis07} H.M.~Wiseman, S.J.~Jones, and A.C.~Doherty, {\it Steering, entanglement, nonlocality, and the EPR paradox}, Phys.~Rev.~Lett.~{\bf 98}, 140402 (2007).
  \bibitem{Qui15} M.T.~Quintino, T.~V\'ertesi, D.~Cavalcanti, R.~Augusiak, M.~Demianowicz, A.~Ac\'{\i}n, and N.~Brunner, {\it Inequivalence of entanglement, steering, and Bell nonlocality for general measurements}, Phys.~Rev.~A {\bf 92}, 032107 (2015).
  \bibitem{Bru14} N.~Brunner, D.~Cavalcanti, S.~Pironio, V.~Scarani, and S.~Wehner, {\it Bell nonlocality}, Rev.~Mod.~Phys.~{\bf 86}, 419 (2014).
  \bibitem{Cav17} D.~Cavalcanti and P.~Skrzypczyk, {\it Quantum steering: a review with focus on semidefinite programming}, Rep.~Prog.~Phys.~{\bf 80}, 024001 (2017).
  \bibitem{Pop95} S.~Popescu, {\it Bell's inequalities and density matrices: revealing ``hidden'' nonlocality}, Phys.~Rev.~Lett.~{\bf 74}, 2619 (1995).
  \bibitem{Gis96} N.~Gisin, {\it Hidden quantum nonlocality revealed by local filters}, Phys.~Lett.~A {\bf 210}, 151-156 (1996).
  \bibitem{Hir13} F.~Hirsch, M.T.~Quintino, J.~Bowles, and N.~Brunner, {\it Genuine hidden quantum nonlocality}, Phys.~Rev.~Lett.~{\bf 111}, 160402 (2013).
  \bibitem{Hir16} F.~Hirsch, M.T.~Quintino, J.~Bowles, T.~V\'ertesi, and N.~Brunner, {\it Entanglement without hidden nonlocality}, New J.~Phys.~{\bf 18}, 113019 (2016).
  \bibitem{Gal14} R.~Gallego, L.~W\"urflinger, R.~Chaves, A.~Acin, and M.~Navascu\'es, {\it Nonlocality in sequential correlation scenarios}, New J.~Phys.~{\bf 16}, 033037 (2014).
  \bibitem{Sil15} R.~Silva, N.~Gisin, Y.~Guryanova, and S.~Popescu, {\it Multiple observers can share the nonlocality of half of an entangled pair by using optimal weak measurements}, Phys.~Rev.~Lett.~{\bf 114}, 250401 (2015).
  \bibitem{Cur17} F.J.~Curchod, M.~Johansson, R.~Augusiak, M.J.~Hoban, P.~Wittek, and A.~Acin, {\it Unbounded randomness certification using sequences of measurements}, Phys.~Rev.~A {\bf 95}, 020102 (2017).
  \bibitem{Tav18} A.~Tavakoli, A.~Cabello, {\it Quantum predictions for an unmeasured system cannot be simulated with a finite-memory classical system}, Phys.~Rev.~A {\bf 97}, 032131 (2018).
  \bibitem{Sch17} M.~Schiavon, L.~Calderaro, M.~Pittaluga, G.~Vallone, and P.~Villoresi, {\it Three-observer Bell inequality violation on a two-qubit entangled state}, Quant.~Sci.~Technol.~{\bf 2}, 015010 (2017).
  \bibitem{Hu16} M.J.~Hu, Z.Y.~Zhou, X.M.~Hu, C.F.~Li, G.C.~Guo, and Y.S.~Zhang, {\it Experimental sharing of nonlocality among multiple observers with one entangled pair via optimal weak measurements}, arXiv:1609.01863.
  \bibitem{Ber18} A.~Bera, S.~Mal, A.S.~De, and U.~Sen, {\it Witnessing entanglement sequentially: maximally entangled states are not special}, Phys.~Rev.~A {\bf 98}, 062304 (2018).
  \bibitem{Sah18} S.~Saha, D.~Das, S.~Sasmal, D.~Sarkar, K.~Mukherjee, A.~Roy, and S.S.~Bhattacharya, {\it Sharing of tripartite nonlocality by multiple observers measuring sequentially at one side}, arXiv:1807.08498 (2018).
  \bibitem{Sas18} S.~Sasmal, D.~Das, S.~Mal, and A.S.~Majumdar, {\it Steering a single system sequentially by multiple observers}, Phys.~Rev.~A {\bf 98}, 012305 (2018).
  \bibitem{Wer89} R.F.~Werner, {\it Quantum states with Einstein-Podolsky-Rosen correlations admitting a hidden-variable model}, Phys.~Rev.~A {\bf 40}, 4277 (1989).
  \bibitem{Bus86} P.~Busch, {\it Unsharp reality and joint measurements for spin observables}, Phys.~Rev.~D {\bf 33}, 2253-2261 (1986).
  \bibitem{Mal16} S.~Mal, A.S.~Majumdar, and D.~Home, {\it Sharing of nonlocality of a single member of an entangled pair is not possible by more than two unbiased observers on the other wing}, Mathematics {\bf 4}, 48 (2016).
  \bibitem{Eme05} J.~Emerson, R.~Alicki, and K.~\.Zyczkowski, {\it Scalable noise estimation with random unitary operators}, J.~Opt.~B: Quant.~Semiclass.~Opt.~{\bf 7}, 347-352 (2005).
  \bibitem{DiM14} O.~Di Matteo, {\it A short introduction to unitary 2-designs}, CS867/QIC890 (2014).
  \bibitem{Dan09} C.~Dankert, R.~Cleve, J.~Emerson, and E.~Livine, {\it Exact and approximate unitary 2-designs: constructions and application}, Phys.~Rev.~A {\bf 80}, 012304 (2009).
  \bibitem{DiV02} D.P.~DiVicenzo, D.W.~Leung, and B.M.~Terhal, {\it Quantum data hiding}, IEEE Trans.~Inf.~Theory Vol.~48 No.~3, 580-599 (2002).
  \bibitem{Kla05} A.~Klappenecker and M.~Roetteler, {\it Mutually unbiased bases are complex projective 2-designs}, Proceedings of the International Symposium on Information Theory, Adelaide, Australia, 1740-1744, (2005).
  \bibitem{Des18} S.~Designolle, P.~Skrzypczyk, F.~Fr\"owis, and N.~Brunner, {\it Quantifying measurement incompatibility of mutually unbiased bases}, Phys.~Rev.~Lett.~{\bf 122}, 050402 (2019).
  \bibitem{Qui14} M.T.~Quintino, T. V\'ertesi, and N. Brunner, {\it Joint Measurability, Einstein-Podolsky-Rosen Steering, and Bell Nonlocality}, Phys.~Rev.~Lett.~{\bf 113}, 160402 (2014).
  \bibitem{Uol14} R.~Uola, T.~Moroder, and O.~G\"uhne, {\it Joint measurability of generalized measurements implies classicality}, Phys.~Rev.~Lett.~{\bf 113}, 160403 (2014).
  \bibitem{Uol15} R.~Uola, C.~Budroni, O.~G\"uhne, J.P.~Pellonp\"a\"a, {\it One-to-one mapping between steering and joint measurability problems}, Phys.~Rev.~Lett.~{\bf 115}, 230402 (2015).
  \bibitem{Hir17} F.~Hirsch, M.T.~Quintino, T.~V\'ertesi, M.~Navascu\'es, and N.~Brunner, {\it Better local hidden variable models for two-qubit Werner states and an upper bound on the Grothendieck constant KG(3)}, Quantum {\bf 1}, 3 (2017).
  \bibitem{Bri16} S.~Brierley, M.~Navascu\'es, and T.~V\'ertesi, {\it Convex separation from convex optimization for large-scale problems}, arXiv:1609.05011.

\end{thebibliography}
\end{document}